\documentclass[10.5pt]{article}
\usepackage[utf8]{inputenc}
\usepackage{geometry}
\usepackage{authblk}
\usepackage{algorithm,algpseudocode,amsfonts,amsmath,amssymb,amsthm}
\usepackage{enumitem,graphicx,hyperref,pgfplots,stmaryrd,subfig,multirow}
\usepackage{booktabs}
\usepackage{CJKutf8,dsfont}
\newlist{todolist}{itemize}{2}
\setlist[todolist]{label=$\square$}
\hypersetup{
colorlinks=true,
citecolor=blue(ryb),
linkcolor=crimson,
urlcolor=cyan
}
\definecolor{azure}{rgb}{0.0, 0.5, 1.0}
\definecolor{frenchblue}{rgb}{0.0, 0.45, 0.73}
\definecolor{forestgreen(traditional)}{rgb}{0.0, 0.27, 0.13}
\definecolor{mygreen}{rgb}{0.09, 0.45, 0.27}
\definecolor{myblue}{rgb}{0.2383,0.5195,0.7734}
\definecolor{mygreen}{rgb}{0.6445,0.9297,0.0039}
\definecolor{darklavender}{rgb}{0.45, 0.31, 0.59}
\definecolor{americanrose}{rgb}{1.0, 0.01, 0.24}
\definecolor{pigblue}{rgb}{0.2, 0.2, 0.6}
\definecolor{blue(ryb)}{rgb}{0.01, 0.28, 1.0}
\definecolor{amethyst}{rgb}{0.6, 0.4, 0.8}
\definecolor{deepmagenta}{rgb}{0.8, 0.0, 0.8}
\definecolor{carminered}{rgb}{1.0, 0.0, 0.22}
\definecolor{iris}{rgb}{0.35, 0.31, 0.81}

\newtheoremstyle{mystyle} 
    {\topsep}                    
    {\topsep}                    
    {\normalfont}                
    {}                           
    {\bfseries}                  
    {.}                          
    {.5em}                       
    {}                           
\theoremstyle{mystyle}

\def \fronorm {\textsf{F}}
\def \inv {{-1}}

\renewcommand{\vec}{\mathrm{vec}}

\newcommand{\reals}{{\mbox{\bf R}}}

\newcommand{\prox}{\operatorname{prox}}

\newcommand{\argmin}{\mathop{\rm arg\,min}}

\newcommand{\sign}{\mathop{\bf sign}}

\def \sign    {\operatorname{sign}}
\def \tr      {\operatorname{tr}}

\renewcommand{\Pr}{\operatorname{Pr}}

\def \reals    {{\mathbb R}}

\def\ccalA{{\ensuremath{\mathcal A}}}
\def\ccalB{{\ensuremath{\mathcal B}}}

\def\ccalX{{\ensuremath{\mathcal X}}}

\def\ccal0{{\ensuremath{\mathcal 0}}}
\def\bbA{{\ensuremath{\boldsymbol A}}}
\def\bbB{{\ensuremath{\boldsymbol B}}}

\def\bbD{{\ensuremath{\boldsymbol D}}}

\def\bbH{{\ensuremath{\boldsymbol H}}}
\def\bbI{{\ensuremath{\boldsymbol I}}}

\def\bbL{{\ensuremath{\boldsymbol L}}}
\def\bbM{{\ensuremath{\boldsymbol M}}}

\def\bbP{{\ensuremath{\boldsymbol P}}}

\def\bbW{{\ensuremath{\boldsymbol W}}}

\def\bbX{{\ensuremath{\boldsymbol X}}}

\def\bba{{\ensuremath{\boldsymbol a}}}
\def\bbb{{\ensuremath{\boldsymbol b}}}

\def\bbr{{\ensuremath{\boldsymbol r}}}

\def\bbu{{\ensuremath{\boldsymbol u}}}
\def\bbv{{\ensuremath{\boldsymbol v}}}
\def\bbs{{\ensuremath{\boldsymbol s}}}

\def\bbx{{\ensuremath{\boldsymbol x}}}
\def\bby{{\ensuremath{\boldsymbol y}}}
\def\bbz{{\ensuremath{\boldsymbol z}}}
\def\bb0{{\ensuremath{\boldsymbol 0}}}
\def\bbbeta{\boldsymbol{\beta}}

\def\bbtheta{\boldsymbol{\theta}}

\def\hbtheta{\hat\bbtheta}

\newcommand\blue[1]{{\color{blue} \textbf{#1}}}

\definecolor{crimson}{rgb}{0.86, 0.08, 0.24}
\definecolor{scarlet}{rgb}{1.0, 0.13, 0.0}
\definecolor{hookersgreen}{rgb}{0.0, 0.44, 0.0}
\definecolor{cultramarine}{rgb}{0.25, 0.0, 1.0}

\newcommand\revised[1]{{\color{black} #1}}

\def \TV {\operatorname{TV}}
\def \unvec {\operatorname{unvec}}
\def \vec {\operatorname{vec}}

\def \BIC {\text{BIC}}

\title{{\sf Generalized Nonnegative Structured Kruskal Tensor Regression}}
\author[1]{Xinjue Wang \thanks{This work is supported by the Research Council of Finland under Grant 359848, and Grant 357715.}}
\affil[1]{Department of Information and Communications Engineering, Aalto University}
\author[1]{Esa Ollila}
\author[1]{Sergiy A. Vorobyov}
\author[2]{Ammar Mian}
\affil[2]{LISTIC, University Savoie Mont Blanc}

\begin{document}

\maketitle

\begin{abstract}
This paper introduces Generalized Nonnegative Structured Kruskal Tensor Regression (NS-KTR), a novel tensor regression framework that enhances interpretability and performance through mode-specific hybrid regularization and nonnegativity constraints. 
Our approach accommodates both linear and logistic regression formulations for diverse response variables while addressing the structural heterogeneity inherent in multidimensional tensor data. 
We integrate fused LASSO, total variation, and ridge regularizers—each tailored to specific tensor modes—and develop an efficient alternating direction method of multipliers (ADMM)-based algorithm for parameter estimation. 
Comprehensive experiments on synthetic signals and real hyperspectral datasets demonstrate that NS-KTR consistently outperforms conventional tensor regression methods.
The framework's ability to preserve distinct structural characteristics across tensor dimensions while ensuring physical interpretability makes it especially suitable for applications in signal processing and hyperspectral image analysis.
The source code is available at~\url{https://github.com/xnnjw/General-NSKTR}.

\end{abstract}


\section{Introduction}
\label{sec:intro}
Tensor decompositions have emerged as powerful analytical tools across diverse fields including signal/image processing~\cite{Sidiropoulos17TensorSPML,cichocki2015tensor,guo2012tensor,lock2018tensor,Long2019Feb,Zhu2025MaySP_TensorSVD}, chemometrics~\cite{yan2019structured}, geophysics~\cite{huang2019reweightedTensorSAR, xu20nolocalCPHSI}, and neuroscience~\cite{Cong15TensorEEG, Huang20TensorBioinfo, zhou13tensor, Chen2022AugTVTensor}. 
Their effectiveness stems from their ability to approximate high-dimensional tensors with low-rank decompositions, offering efficient dimensionality reduction while preserving essential structural information.
For example, tensor decomposition techniques~\cite{carroll1970analysis,harshman1970foundations,tucker1963implications} are applied in hyperspectral image (HSI) analysis to extract low-rank structures for dimensionality reduction~\cite{xu20nolocalCPHSI}, and used in electroencephalogram (EEG) analysis to capture latent patterns across multiple dimensions~\cite{Cong15TensorEEG}.
Over the past decade, tensor regression (TR) models have received attention, with numerous approaches proposed in the literature, including Tucker tensor regression~\cite{li2018tucker}, low-rank orthogonally decomposable tensor regression~\cite{poythress2021low}, Bayesian Kruskal tensor regression (KTR)~\cite{guhaniyogi2017bayesian}, Bayesian low rank tensor ring completion~\cite{Long21BayesianTensorRing}, graph-regularized tensor regression~\cite{xu2023graphtenreg}, and tensor regression network~\cite{Kossaifi20TRNetworks}.

\subsection{Motivation}
The Kruskal tensor regression (KTR) model~\cite{Kruskal1977kruskal} assumes a rank-$R$ {CANDECOMP/PARAFAC} decomposition (CPD)~\cite{carroll1970analysis,harshman1970foundations} for the tensor regression parameter $\ccalB$, thus reducing the number of unknown variables.  
However, real-world tensor data, such as HSIs or brain connectivity networks—often exhibit multi-dimensional structural heterogeneity (e.g., smoothness, sparsity, or anatomical connectivity patterns) that demands not only dimensionality reduction but also targeted regularization to preserve these structural characteristics. 
Furthermore, classical Frobenius norm-based tensor decompositions struggle with non-Gaussian measurements (e.g., binary data), motivating the need for generalized tensor regression frameworks that integrate adaptive regularization and flexible regression losses to ensure interpretability, structural fidelity, and compatibility with diverse response variables.

A critical observation driving our approach is that, in many applications, both the tensor covariates and the regression targets are inherently non-negative. 
For instance, in HSI analysis, the regression coefficient $(\ccalB)_{ijk}$ corresponding to the $(i,j,k)$th feature of the 3D-covariate tensor $\ccalX$ not only indicates the presence/absence of that image signature but also quantifies its intensity. 
This physical interpretation naturally motivates nonnegativity constraints on the elements of $\ccalB$. 
Theoretical justification for this constraint comes from~\cite{lim2005nonnegative, lim2009nonnegative}, which demonstrates that nonnegativity constraints ensure the existence of global solutions for CPD approximation problems, improving both optimization stability and result interpretability.
Beyond nonnegativity, tensor image covariates exhibit rich structural characteristics that should be reflected in the tensor parameter $\ccalB$. 
HSIs, for example, typically display smooth variations across spectral bands while maintaining locally constant profiles in spatial dimensions.
These distinct structural properties across different tensor modes call for a targeted, dimension-specific regularization approach rather than uniform regularization across all dimensions.

\subsection{Problem to be addressed}
We therefore propose a hybrid regularization framework that combines fused LASSO~\cite{Tibshirani2005FusedLASSO} (promoting piecewise constancy) and ridge regularizers (promoting smoothness) on different dimensions of the tensor parameter.
For a vector $\bbu\in\reals^m$, our hybrid regularizer takes the form:
\begin{align} \label{eq:intro_nfl}
    h(\bbu) = \lambda_1 \|\bbu\|_1 + \lambda_2 \|\bbu\|_{\TV} + \frac{\lambda_3}{2} \|\bbu\|_2^2 + \iota_{\geq0}(\bbu),
\end{align}
where $(\lambda_1, \lambda_2, \lambda_3)$ are nonnegative penalty parameters controlling the weights of different property preferences, and $   \iota_{\geq 0}(\bbu) $ is the indicator function that enforces nonnegativity, i.e., it equals $0$ if $\bbu \geq 0 $ and $+\infty$ otherwise. 
The latter term can be optional, only put in effect when nonnegativity of the parameter is desired. 
By adjusting the parameters across different tensor modes, our model can adapt to the specific structural characteristics of each dimension.
    
Another crucial challenge in tensor analysis concerns the nature of the data itself. 
Tensor datasets frequently consist of non-Gaussian data types, such as binary, categorical, or count-based measurements~\cite{hu2022generalized}. 
For example, brain connectivity networks~\cite{zhou13tensor} represent interactions through binary adjacency matrices, and HSI datasets often involve categorical pixel labels~\cite{guo2016SVM_tensor}. 
Standard tensor decomposition methods typically minimize the Frobenius norm between the decomposition model and the data, implicitly assuming Gaussian noise. 
This assumption becomes problematic for binary valued data, often resulting in poor predictive performance and limited interpretability.

The alternating direction method of multipliers (ADMM)~\cite{gabay1976dual, combettes2011proximal, boyd2011distributed, Wang2024SepADMMTensor} is a powerful tool in tensor-related optimization, applied in tasks like tensor restoration and regression~\cite{huang2016flexible, chen2020robust, wang2023guaranteed,lu2020high,Wang2022JunADMMTV}. 
ADMM excels in problems with separable objectives and constraints, making it ideal for our framework, which features a non-convex loss function, nonnegativity constraints, and structured regularization. 

\revised{
\subsection{Related Work}
A number of tensor regression models have been proposed to address the challenge of high-dimensional tensor-valued covariates. 
Tensor linear regression (TLR)~\cite{zhou13tensor} provides a basic Kruskal-based regression framework without specialized regularization, while Tucker tensor regression (TuckerTR)~\cite{li2018tucker_comp2} employs a flexible Tucker decomposition with a core tensor. 
Extensions of quantile regression to tensor settings have been studied in both Tucker and CPD formulations~\cite{lu2020high, li2021tensorquantile} for improved robustness.
These models form the basic paradigms of tensor regression and serve as important references for positioning our contribution.
In~\cite{li2021tensorquantile}, a CPD-based quantile tensor regression (TQtR) was developed with fusion and smoothness penalties, optimized via block relaxation and ADMM, and demonstrated interpretability in neuroimaging. 
In~\cite{lu2020high}, a Tucker-based quantile tensor regression was proposed in both sparse and dense variants, with $\ell_1$ regularization on factor matrices, convergence guarantees, and ADMM-based optimization, applied to video data. 
In~\cite{xu2023graphtenreg}, graph Laplacian regularization was introduced to capture domain structures in tensor regression.
In~\cite{ouyang2020tensorgenomic}, Tucker-based regression with direct sparsity-inducing penalties was investigated in genomic applications. 
Beyond regression, \cite{hu2022generalized} incorporated auxiliary covariates into tensor factorization, and \cite{durand2024NNTensorTSP} proposed smoothness-enforcing penalties under nonnegativity constraints.
Compared with these works, our proposed framework builds on KTR and uniquely unifies mode-specific hybrid regularization with nonnegativity within generalized regression models.
}

\subsection{Main Contributions}
In this work, we propose a novel tensor regression framework based on the KTR model, with three key contributions:
\begin{enumerate}[noitemsep,topsep=0pt,parsep=0pt,partopsep=0pt]
    \item We introduce a nonnegative structured Kruskal tensor regression (NS-KTR) approach that integrates nonnegativity constraints with mode-specific hybrid regularization. 
    This method adaptively promotes sparsity, piecewise constancy, and smoothness, tailored to the inherent structural characteristics of each tensor dimension.
    \item Our framework extends beyond Gaussian response assumption by incorporating both linear and logistic tensor regression variants. This enables effective modeling of continuous and binary response types through a unified computational approach.
    \item We develop an efficient alternating optimization algorithm using the ADMM. 
    This algorithm demonstrates superior performance across synthetic and real-world datasets, with comprehensive experiments showing consistent improvements in estimation accuracy for various signal types and performance gains in HSI data.
\end{enumerate}

\subsection{Paper Organization}
The remainder of this paper is organized as follows. 
Section~\ref{sec:preliminaries} provides preliminary background on tensor notation, decompositions, and regression models.
Section~\ref{sec:NSKTR_proposed} introduces our proposed NS-KTR method, detailing the hybrid regularization strategy, mode-specific structural constraints, and extension to generalized regression models. 
We then develop an alternating optimization algorithm based on ADMM to solve the resulting estimation problems. 
Section~\ref{sec:num_exp} presents numerical experiments on both synthetic and real hyperspectral datasets, demonstrating the superior performance of our method across various signal types, sample sizes, and noise conditions.
Section~\ref{sec:conclusion} concludes the paper and discusses the future work.

\section{Preliminaries}
\label{sec:preliminaries}
We use $\ccalB = (b_{i_1 \cdots i_D})$ to denote a $D$-way tensor of size $I_1 \times \cdots \times I_D$. 
The mode-$d$ matricization~\cite{kolda2009tensor} of $\ccalB$ is defined as a matrix $\bbB_{(d)} \in\reals^{I_d \times \prod_{d' \ne d} I_{d'}}$, which reshapes tensor $\ccalB$ to a $I_d \times \prod_{d' \ne d} I_{d'}$ matrix such that the tensor's $(i_1,\ldots,i_D)$ element corresponds to the $(i_d,j)$ element of the matrix $\bbB_{(d)}$, where
$
    j = 1 + \sum_{d'\ne d} (i_{d'}-1) \prod_{d''<d',d'' \ne d} I_{d''}.
$
The vectorization operator $\mathrm{vec}(\cdot)$ maps a tensor into a vector by stacking the columns of $\bbB_{(d)}$ on top of each other.
The tensor inner product between two tensors of same size are also denoted by the inner product of their vectorized or mode-$d$ matricized counterparts as 
$\langle \ccalA, \ccalB \rangle =\langle \mathrm{vec}(\ccalA), \mathrm{vec}(\ccalB) \rangle  = \langle \bbA_{(d)} , \bbB_{(d)} \rangle$,
where the latter inner product for matrices can also be expressed compactly using matrix trace as $ \langle \bbA_{(d)} , \bbB_{(d)} \rangle = \mathrm{tr}(\bbA_{(d)} \bbB_{(d)}^\top )$.
%

The rank-$R$ CPD expresses a tensor as a linear combination of rank-$1$ tensors: 
\begin{align} 
\label{eq:CPD}
\ccalB & \equiv \llbracket \bbB_{1}, \ldots, \bbB_{D}  \rrbracket  = \sum_{r=1}^R \bbB_1(\cdot, r)  \circ \cdots \circ \bbB_D(\cdot, r),
\end{align}
where $\bbB_d \in \reals^{I_d \times R}$ for $d=1,\ldots,D$ are latent factor matrices, $\bbB_d(\cdot,r) \in \reals^{I_d}$ denotes the $r$-th column of matrix, and $\circ$ denotes the outer product.
A tensor admitting decomposition~\eqref{eq:CPD} is also referred to as a Kruskal tensor \cite{kolda2009tensor}. 
Consider two matrices $\bbA = (\bba_1 \,  \cdots \, \bba_n )\in \reals^{m \times n}$ and $\bbB = (\bbb_1 \cdots \bbb_q)  \in \reals^{p \times q}$. 
If $\bbA$ and $\bbB$ have the same number of columns $n=q$, then the Khatri-Rao product is defined as a columnwise Kronecker product
\begin{align}
    \bbA \odot \bbB = (
    \boldsymbol{a}_1 \otimes \boldsymbol{b}_1  \ \boldsymbol{a}_2 \otimes \boldsymbol{b}_2 \ \ldots \ \boldsymbol{a}_n \otimes \boldsymbol{b}_n),
\end{align}
where $\otimes$ denotes the Kronecker product.
If $\ccalB \in \reals^{I_1 \times \cdots \times I_D}$ is a rank-$R$  Kruskal tensor~\eqref{eq:CPD}, then the inner-product between $\ccalB$ and tensor $\ccalX \in \reals^{I_1 \times \cdots \times I_D}$ can be found as
\begin{align}
        \langle \ccalX, \ccalB \rangle 
        & = \langle \bbX_{(d)} \bbB_{(-d)}, \bbB_d  \rangle =  \tr ((\bbX_{(d)} \bbB_{(-d)})^\top \bbB_d ) \notag \\
        & = \vec(\bbX_{(d)} \bbB_{(-d)})^\top \vec(\bbB_d), \label{eq:tensor_innerprod}
\end{align}
where $\bbB_{(-d)}$ is a Khatri-Rao product between all $d'$-th factor matrices ($d' = 1,\ldots, D, d'\not=d$) of $\ccalB$
\begin{align}
    \label{eq:KTR_B}
    \bbB_{(-d)} & = \bbB_D \odot \cdots \odot \bbB_{d+1} \odot \bbB_{d-1} \odot \cdots \odot \bbB_1,
\end{align}
and $\bbX_{(d)} \bbB_{(-d)}$ is the mode-$d$ matricized tensor times Khatri-Rao product (MTTKRP), which is the key computational element in CP decomposition~\cite{kolda2009tensor}.
We use 
\begin{align} 
    \label{eq:vecandmat}
    \bbbeta_d = \vec(\bbB_d) \in \reals^{I_dR}
\end{align}
to denote the vectorized form of the $d$th factor matrix, with the inverse operation
\begin{align} 
    \label{eq:unvecop}
    \bbB_d = \unvec(\bbbeta_d) \in \reals^{I_d\times R}.
\end{align}
Additionally, we use $\bbbeta_{dr} = \bbB_{d}(\cdot,r) \in \reals^{I_d}$ to denote the $r$th column of the $d$th factor matrix, and thus, the vectorized $\bbbeta_d$ concatenates all columns of~$\bbB_{d}$.

\subsection{Linear Regression Model}
\label{sec:sub_linearregmodel}
In tensor linear regression, a scalar response $y_i\in\reals$ is modeled as a linear function of a tensor covariate $\ccalX_i\in\reals^{I_1 \times \cdots \times I_D}$
\begin{align*}
    y_i = \langle \ccalX_i, \ccalB\rangle + e_i, \ i=1,\ldots,N,
\end{align*}
where $\ccalB\in\reals^{I_1 \times \cdots \times I_D}$ is coefficient tensor, and $e_i, i=1,\ldots,N$ are independent and identically distributed (i.i.d.) errors following normal distribution. 
To estimate $\ccalB$, the optimization problem can be formulated as
\begin{align*}
    \min_\ccalB \frac{1}{2} \sum_{i=1}^N(y_i - \langle \ccalX_i, \ccalB\rangle)^2 + h(\ccalB),
\end{align*}
where $h(\cdot)$ is a regularization term that imposes structural constraints or any prior knowledge on $\ccalB$. 
Specific forms of $h(\cdot)$ will be detailed in Section~\ref{sec:NSKTR_proposed}.

\subsection{Logistic Regression Model}
\label{sec:sub_logisticregmodel}
For tensor logistic regression with binary responses \(y_i \in \{-1, 1\}\), the probability of the positive class is modeled as
\begin{align*}
    \Pr(y_i=1) = \frac{1}{1 + \exp(-\langle \mathcal{X}_i, \mathcal{B} \rangle)},
\end{align*}
where \(\mathcal{X}_i\) and \(\mathcal{B} \in \mathbb{R}^{I_1 \times \cdots \times I_D}\) are the tensor covariate and coefficient tensor, respectively.
The estimation of \(\mathcal{B}\) involves minimizing the negative log-likelihood with a regularization term
\begin{align} \label{eq:logistic_objective}
    \min_\ccalB \sum_{i=1}^N\log(1+\exp(-y_i \langle \ccalX_i, \ccalB \rangle)) + h(\ccalB), 
\end{align}
where the first term is the summation of binomial deviance losses of the data points, and \(h(\mathcal{B})\) is a regularization term to be specified in Section~\ref{sec:NSKTR_proposed}. 

\section{Nonnegative Structured Kruskal Tensor Regression}
\label{sec:NSKTR_proposed}
In tensor regression, a critical challenge is the exponential growth of parameters. 
For a coefficient tensor $\mathcal{B}$ of size $I_1 \times \cdots \times I_D$, the total number of parameters is $\prod_{d=1}^D I_d$, which quickly exceeds the available sample size $N$ as dimensions increase.
By assuming a low-rank structure for $\mathcal{B}$ through a rank-$R$ CP decomposition, we reduce the number of parameters to $R\sum_{d=1}^D I_d$, substantially mitigating overfitting risk while preserving computational efficiency.
The CPD of tensor $\mathcal{B}$ with rank-$R$, where $R\in \mathbb{N}^+_0$, allows us to separate the $d$-th factor $\bbB_d$ from the tensor inner product
\begin{align} \label{eq:GKTR_model}
    \langle \ccalX_i, \ccalB\rangle = \langle\ccalX_i, \llbracket \bbB_1, \ldots, \bbB_D \rrbracket \rangle = \vec(\bbX_{i(d)} \bbB_{(-d)})^\top \bbbeta_d,
\end{align}
where $\boldsymbol{\beta}_d = \operatorname{vec}(\boldsymbol{B}_d) \in \mathbb{R}^{I_dR}$ is the vectorized $d$-th factor matrix.

\subsection{Structured Regularization} 
Real-world tensor data often exhibit distinctive structural characteristics across their various dimensions. 
For instance, in HSI, the spectral dimension typically shows smooth variations due to the continuous nature of spectral signatures, while the spatial dimensions are characterized by locally constant regions with meaningful boundaries, reflecting spatial coherence.
To effectively capture these multi-dimensional structures, we propose a hybrid regularization framework that extends the method of~\cite{mairal2010online} by incorporating nonnegativity constraints, explicitly tailoring regularization to the multi-dimensional structural properties of each mode~$d$.

Our overall regularization function $h(\cdot)$ for the joint optimization problem is composed of separate mode-specific regularizers:
\begin{align}
    h(\boldsymbol{\beta}_1, \ldots, \boldsymbol{\beta}_D) = \sum_{d=1}^D h_d(\boldsymbol{\beta}_d),
\end{align}
where each $h_d(\cdot)$ acts independently on its corresponding factor vector $\boldsymbol{\beta}_d$. 
For each mode $d$, we define a general form of regularization function as
\begin{align} \label{eq:h_hybrid}
    h_d(\boldsymbol{\beta}_d) 
    \triangleq \lambda_{d1} \|\boldsymbol{\beta}_d\|_1 + \lambda_{d2} \|\bbD_d \boldsymbol{\beta}_d\|_1 + \frac{\lambda_{d3}}{2} \|\boldsymbol{\beta}_d\|_2^2 + \iota_{\geq 0}(\boldsymbol{\beta}_d),
\end{align}
where $\boldsymbol{\beta}_d = \vec(\bbB_d) \in \mathbb{R}^{I_dR}$ is the vectorized factor matrix, and ($\lambda_{d1}$, $\lambda_{d2}$ $\lambda_{d3}$) are non-negative regularization parameters that control different structural properties, and $\bbD_d \in \mathbb{R}^{(I_d-1)R \times I_dR}$ is the first-order difference matrix defined~as
\begin{align} \label{eq:Dd_difference}
    \bbD_d \triangleq \bbI_R \otimes
    \begin{bmatrix}
                -1 	&  1 		&  		0 & \ldots  &   	0&  0\\
                0 & -1 			& 1 		&\ldots  &  0	&   0\\  
                &  &   	&\vdots  &   &    \\  
                0&   0 		&  		0& \ldots & -1 		&  1
    \end{bmatrix}_{(I_d-1)\times I_d}.
\end{align}
Each component of~\eqref{eq:h_hybrid} serves a distinct purpose.
\textit{(i)} The $\ell_1$ norm term $\|\boldsymbol{\beta}_d\|_1$ promotes sparsity, effectively performing feature selection within each mode.
\textit{(ii)} The total variation term $\|\bbD_d\boldsymbol{\beta}_d\|_1$ encourages piecewise constant patterns, preserving sharp transitions between regions, as introduced in~\cite{rudin1992TV}. 
\textit{(iii)} The quadratic term $\|\boldsymbol{\beta}_d\|_2^2$ induces smoothness, mitigating noise and ensuring stability in the estimation~\cite{mairal2010online}.
\textit{(iv)} The indicator function $\iota_{\geq 0}(\boldsymbol{\beta}_d)$ enforces nonnegativity constraint, which is critical for applications like HSI where physical quantities (e.g., reflectance) are inherently non-negative.

\begin{figure}[!t]
    \centering
    \centerline{\includegraphics[width = 0.85\linewidth]{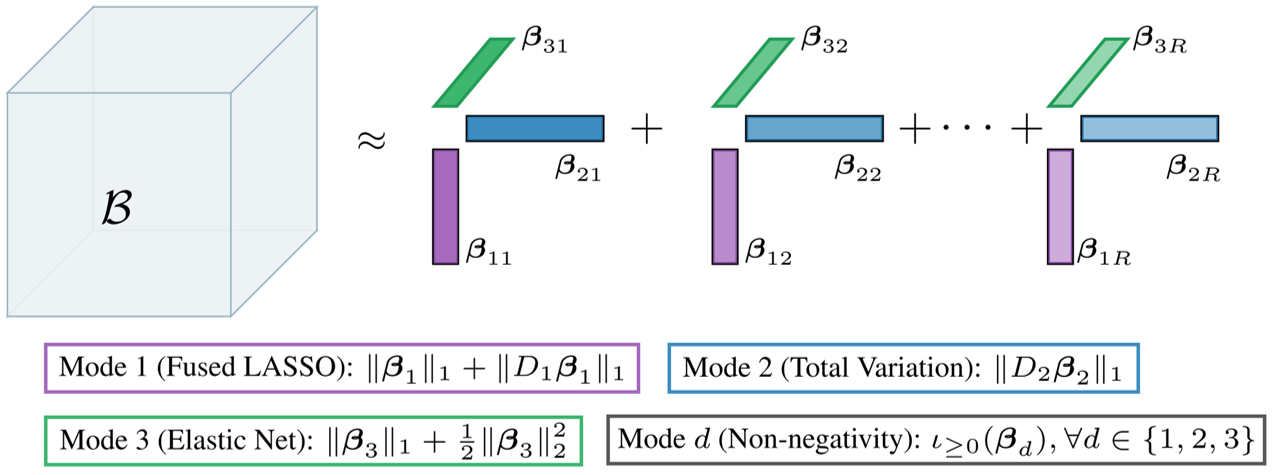}}
    \caption{An example of nonnegative structured CP decomposition. 
    The tensor $\ccalB$ is decomposed into the sum of $R$ rank-1 tensors, each formed by the outer product of nonnegative factor vectors with tailored regularization across different dimensions.
    }
    \label{fig:ExampleCPD}
\end{figure}

By strategically configuring the regularization parameters~\((\lambda_{d1}, \lambda_{d2}, \lambda_{d3})\) and selectively enforcing the nonnegativity constraint $\iota_{\geq 0}$, our proposed hybrid framework enables flexible, mode-specific regularization tailored to the structural characteristics of each tensor mode.
Specifically, several well-established regularization terms can be replicated as follows.
\begin{enumerate}[noitemsep,topsep=0pt,parsep=0pt,partopsep=0pt]
    \item LASSO (Sparsity): When \(\lambda_{d1} > 0\), \(\lambda_{d2} = 0\), \(\lambda_{d3} = 0\), the penalty reduces to the LASSO, promoting sparsity in the factor matrix.
    With the nonnegativity constraint \( \iota_{\geq 0} \cdot \), this becomes nonnegative LASSO, ensuring sparse, non-negative factors.
    \item Total Variation (TV): When \(\lambda_{d1} = 0\), \(\lambda_{d2} > 0\), \(\lambda_{d3} = 0\), the penalty corresponds to TV regularization, encouraging piecewise constant solutions. 
    \item Fused LASSO (Sparsity + Piecewise Constant): When \(\lambda_{d1} > 0\), \(\lambda_{d2} > 0\), \(\lambda_{d3} = 0\), the penalty becomes the fused LASSO, simultaneously enforcing sparsity and piecewise constant behavior. 
    \item Elastic Net (Sparsity + Smoothness): When \(\lambda_{d1} > 0\), \(\lambda_{d2} = 0\), \(\lambda_{d3} > 0\), the penalty aligns with the Elastic Net, balancing sparsity and smoothness.
\end{enumerate}
Fig.~\ref{fig:ExampleCPD} illustrates the nonnegative structured decomposition where a tensor is factorized into rank-1 components with mode-specific regularization properties. 
Each factor matrix preserves distinct structural characteristics while maintaining nonnegativity constraints, capturing the heterogeneous patterns across different tensor dimensions.

By integrating these mode-specific penalties into the NS-KTR framework, we can adapt the regularization to the unique characteristics of each tensor dimension. 
For example, in HSI, the spectral mode may prioritize smoothness (higher \(\lambda_{d3}\)), while the spatial modes emphasize piecewise constant behavior (higher \(\lambda_{d2}\)). 
This adaptability enhances both the interpretability of the factor matrices and the predictive performance of the model, making it particularly suitable for applications processing tensor data with heterogeneous structural properties across dimensions.

\subsection{Alternating Optimization Framework}
\label{sec:sub_AOframework}
After establishing the structured regularization approach, we now formulate the complete Nonnegative Structured Kruskal Tensor Regression (NS-KTR) optimization problem:
\begin{equation} \label{eq:opt_problem_general}
    \min_{ \{\bbbeta_d\}_{d=1}^D } f(\bbbeta_1, \ldots, \bbbeta_D) + h(\bbbeta_1, \ldots, \bbbeta_D),
\end{equation}
where $f(\cdot)$ represents the data fidelity term that varies based on the regression model. 
For linear regression, this term takes the form
\begin{align}
    f(\boldsymbol{\beta}_1, \ldots, \boldsymbol{\beta}_D) = \frac{1}{2}\sum_{i=1}^N
    \left(y_i - \langle\mathcal{X}_i, \llbracket\bbB_1, \ldots, \bbB_D\rrbracket\rangle\right)^2,
\end{align}
while for logistic regression, it becomes
\begin{align}
    f(\boldsymbol{\beta}_1, \ldots, \boldsymbol{\beta}_D) 
    = \sum_{i=1}^N \log(1 + \exp(-y_i\langle\mathcal{X}_i, \llbracket \bbB_1, \ldots, \bbB_D\rrbracket\rangle)).
\end{align}
This optimization problem~\eqref{eq:opt_problem_general} presents two key challenges: it is non-convex with respect to all factor vectors jointly, and the regularization terms along with nonnegativity constraints add further complexity. 
However, to address it, we can exploit the problem's structure.
Particularly, when all but one factor vector are fixed, the problem becomes convex with respect to the remaining factor vectors.
We therefore can build an alternating optimization strategy, updating each factor vector sequentially while keeping others fixed. 
At the $(t+1)$th iteration, we update the $d$th factor vector as
\begin{align} \label{eq:altopt_betad_general1}
    \bbbeta_d^{t+1} = 
    \argmin_{\bbbeta_{d}} f(\bbbeta_{1}^{t+1}, \ldots, \bbbeta_{d-1}^{t+1}, \bbbeta_d, \bbbeta_{d+1}^{t}, \ldots, \bbbeta_D^{t}) + h_d(\bbbeta_d).
\end{align}
Since each mode-specific subproblem in~\eqref{eq:altopt_betad_general1} has the same form, we can use compact notation. 
Define $\bbx = \boldsymbol{\beta}_d$ for the current mode $d \in \{1,\ldots,D\}$, so each subproblem becomes
\begin{align} \label{eq:altopt_betad_general}
    \bbx = \argmin_{\bbx} g(\bbx) + h_d(\bbx),
\end{align}
where  $g(\cdot)$ takes either the linear regression form
\begin{align}
    g(\bbx) &= \frac12 \|\bby - \bbA\bbx\|_2^2, \label{eq:fx_lin}
\end{align}
or the logistic regression form
\begin{align}
    g(\bbx) &= \sum_{i=1}^N \log(1 + \exp(-y_i(\bbA\bbx)_i). \label{eq:fx_log}
\end{align}
Here, matrix $\bbA\in\reals^{N \times I_dR}$ represents the concatenation of vectorized MTTKRP operations across all samples
\begin{align}\label{eq:subsolver_A_MTTKRP}
    \bbA = \begin{pmatrix} \vec(\bbX_{1(d)}\bbB_{(-d)}) & \cdots &\vec(\bbX_{N(d)}\bbB_{(-d)})\end{pmatrix}^\top,
\end{align}
and $\bba_i = \vec(\bbX_{i(d)}\bbB_{(-d)})\in \reals^{I_d R}$ denotes the vectorized MTTKRP for the $i$th sample. 

The complete alternating optimization procedure is outlined in Algorithm~\ref{alg:NS-KTR}. 
This approach effectively decomposes the original high-dimensional non-convex problem into a sequence of manageable convex subproblems. 
However, each subproblem still involves complex regularization terms and constraints that require specialized solvers. 
In the next section, we introduce an efficient ADMM algorithm to solve these subproblems.

\begin{algorithm}[!h]
\caption{\textsf{NS-KTR}: Nonnegative structured  Kruskal tensor regression}
\label{alg:NS-KTR}
\begin{algorithmic}[1]
\State \textbf{Input:} Response $\bby$, tensor covariates $\ccalX_i, \forall i$, rank $R$.
\State \textbf{Initialization:} $\forall d = 1,
\ldots,D$, Randomize or warm start $\bbB_d^{0}$, 
\For{$t=0,1,\ldots, t_{\mathrm{iter}}$}  
    \State $ \bbB^{t+1}_{(-d)} \gets \bbB^{t}_D \odot \cdots \odot \bbB^{t}_{d+1} \odot \bbB^{t}_{d-1} \odot \cdots \odot \bbB^{t}_1, \forall d \in \{ 1, \ldots, D \} $ 
    \For{$d=1, \ldots, D$}  
        \State $ \bbA \gets \begin{pmatrix} \vec(\bbX_{1(d)}\bbB^{t+1}_{(-d)}) & \cdots & \vec(\bbX_{N(d)}\bbB^{t+1}_{(-d)})\end{pmatrix}^\top $ 
        \State $\bbbeta_d^{t+1} \gets \bbx^{t+1}$ by solving~\eqref{eq:altopt_betad_general} with ADMM (Algorithm~\ref{alg:admm_nnfl})
        \State $\bbB_d^{t+1} \gets \unvec(\bbbeta_d^{t+1})$
    \EndFor
\EndFor
\State \textbf{Return:} $(\bbB_1^{t+1},\ldots,\bbB_D^{t+1})$
\end{algorithmic}
\end{algorithm}

\subsection{Proposed ADMM}

To efficiently solve each subproblem in our alternating optimization framework, we develop an ADMM approach. 
ADMM is well-suited for our scenario as it effectively handles the composite structure of our objective with multiple regularization terms and constraints.

We start by reformulating the $d$-th subproblem~\eqref{eq:altopt_betad_general} by introducing auxiliary variables that separate the objective into more manageable components~as
\begin{subequations}
    \label{eq:nnFL}
    \begin{align}
    \min_{\bbx}  & \ g(\bbx) + \lambda_1 \|\bbz_1\|_1 + \lambda_2 \|\bbz_2\|_1 + \frac{\lambda_3}{2}\|\bbx\|_2^2\\
    \textrm{s.t.} & \ \bbx \geq 0, \\
    & \ \bbx - \bbz_1=0, \\
    & \ \bbD_d\bbx- \bbz_2 = 0,
    \end{align}
\end{subequations} 
where $\bbz_1 \in \reals^{I_dR}$ and $\bbz_2 \in \reals^{(I_d-1)R}$ are auxiliary variables that decouple the non-differentiable regularization terms from the data fidelity term.

The augmented Lagrangian for problem~\eqref{eq:nnFL} is then given as
\begin{equation}
\begin{split}
    L(\bbx, \bbz_1, \bbz_2, \bbu_1, \bbu_2) & 
    = g(\bbx) + \lambda_1 \|\bbz_1\|_1 + \lambda_2 \|\bbz_2\|_1  + \bbu_1^T(\bbx-\bbz_1) + \frac\rho2\|\bbx-\bbz_1\|_2^2 \\
    & + \bbu_2^T(\bbD_d\bbx-\bbz_2) + \frac\rho2\|\bbD_d\bbx-\bbz_2\|_2^2 + \frac{\lambda_3}{2}\|\bbx\|_2^2 + \iota_{\geq 0}(\bbx),
\end{split}
\end{equation}
where $\bbu_1 \in \reals^{I_dR}$ and $\bbu_2 \in \reals^{(I_d-1)R}$ are dual variables, and $\rho > 0$ is the penalty parameter that controls the balance between primal feasibility and dual progress.

The ADMM algorithm proceeds by sequentially updating the primal variables $\boldsymbol{x}$, $\boldsymbol{z}_1$, and $\boldsymbol{z}_2$, followed by updating dual variables $\boldsymbol{u}_1$ and $\boldsymbol{u}_2$.
\\
\noindent \underline{1. $\bbx$-update (Primal Update)}
\begin{align}
    \label{eq:admm_primal_compact}
    \bbx^{k+1} = \argmin_{\bbx \geq 0}L(\bbx, \bbz_1^k, \bbz_2^k, \bbu_1^k, \bbu_2^k).
\end{align}

For {\it linear regression}, where $g(\bbx) = \frac12 \|\bby - \bbA\bbx\|_2^2$, by expanding the right hand side (RHS) of~\eqref{eq:admm_primal_compact}, and transforming it to a quadratic form, we have
\begin{align}
    \bbx^{k+1} = \argmin_{\bbx\geq 0} L_\bbx,
\end{align}
where 
\begin{equation}
    \begin{split}
        L_{\bbx} \triangleq & \dfrac12 \left\|\bby - \bbA\bbx\right\|_2^2 + \frac{\lambda_3}{2}\|\bbx\|_2^2 + \dfrac\rho2\left\|\bbx-\bbz_1^k+\frac{\bbu_1^k}{\rho}\right\|_2^2 + \dfrac\rho2\left\|\bbD_d\bbx-\bbz_2^k+\frac{\bbu_2^k}{\rho}\right\|_2^2 
    \end{split}
\end{equation}
collects the terms of $L(\bbx, \bbz_1, \bbz_2, \bbu_1, \bbu_2)$ that depend on $\bbx$.
The first derivative of $L_\bbx$ is
\begin{equation}
    \begin{split}
        \dfrac{\partial L_\bbx}{\partial \bbx}
        = & \bbA^\top\bbA\bbx - \bbA^\top \bby + \lambda_3 \bbx + \rho \bbx - \rho \bbz^k_1 + \bbu^k_1 + \bbD_d^\top \bbu^k_2 \\
        & + \rho \bbD_d^\top \bbD_d\bbx - \rho \bbD_d^\top \bbz_2^k.
    \end{split}
\end{equation}
Equating $\dfrac{\partial L_\bbx}{\partial \bbx}$ to 0, we obtain
\begin{align}
    \label{eq_admm:gradient}
    \bbx  = \left[ \bbM^\inv \bbb^k \right]_+,
\end{align}
where 
\begin{align}
    \bbM & = \bbA^\top\bbA + (\rho + \lambda_3) \bbI + \rho \bbD_d^\top\bbD_d, \\
    \bbb^k & = \bbA^\top\bby + \rho (\bbz_1^k - \bbu_1^k/\rho) + \rho \bbD_d^\top (\bbz_2^k - \bbu_2^k/\rho),
\end{align}
and $\left[ \cdot \right]_+$ represents element-wise projection onto the non-negative orthant.
Note that $\bbM$ is symmetric and positive definite 
\footnote{The inverse of $\bbM$ can be efficiently solved with backslash operator in MATLAB.
We compute the inverse of $\bbM$ once and reuse it in the following iterations.}.
However, this does not waive the heavy computation of inverse of matrix $\bbM$ when its size $N$ is large.

For {\it logistic regression}, where $g(\bbx) = \sum_{i=1}^N \log(1 + \exp(-y_i(\bbA\bbx)_i)$, we aim to solve
\begin{align}
    \bbx^{k+1} = \argmin_{\bbx \geq 0} L_\bbx,
\end{align}
where 
\begin{equation}
    \begin{split}
        L_\bbx \triangleq & \sum_{i=1}^N \log\big( 1 + \exp(-y_i(\bbA^\top\bbx)_i) \big) + \frac{\lambda_3}{2}\|\bbx\|_2^2  + \dfrac\rho2\left\|\bbx-\bbz_1^k+\frac{\bbu_1^k}{\rho}\right\|_2^2 \\
        & + \dfrac\rho2\left\|\bbD\bbx-\bbz_2^k+\frac{\bbu_2^k}{\rho}\right\|_2^2 
    \end{split}
\end{equation}
collects the terms of $L(\bbx, \bbz_1, \bbz_2, \bbu_1, \bbu_2)$ that depend on~$\bbx$.
Since this subproblem lacks a closed-form solution, we employ Newton's method with line search, as detailed in Algorithm~\ref{alg:newtonforadmm}
\footnote{Note that in the line~\ref{eq:alg_hessian} of Algorithm~\ref{alg:newtonforadmm}, the inverse of the Hessian matrix $\bbH_\bbx^{-1}$ can be computed efficiently using the matrix inversion lemma (Sherman-Morrison-Woodbury formula). 
By restructuring $\bbH_x = \bbA^\top \bbW \bbA + (\rho + \lambda_3)\bbI + \rho \bbD_d^\top \bbD_d$, where $\bbW$ is a diagonal matrix with entries $\bbW_{ii} = {\exp(y_i(\bba_i^\top \bbx))}/{(1+\exp(y_i(\bba_i^\top \bbx)))^2}$, and defining $\bbP = (\rho + \lambda_3)\bbI + \rho \bbD_d^\top \bbD_d$, we obtain $\bbH_\bbx^{-1} = \bbP^{-1} - \bbP^{-1}\bbA^\top(\bbW^{-1} + \bbA\bbP^{-1}\bbA^\top)^{-1}\bbA\bbP^{-1}$. 
The matrix $\bbP$ has a structured form whose inverse can be efficiently computed, avoiding direct inversion of the full Hessian.}.
The Newton's method efficiently handles the smooth but nonlinear objective by iteratively approximating it with quadratic models based on gradient and Hessian information.

\begin{algorithm}[!h]
\caption{Newton's method for primal update in ADMM (Algorithm~\ref{alg:admm_nnfl})}
\begin{algorithmic}[1]
\State \textbf{Input}: Response $\{y_i\}_{i=1}^N$ where $y_i\in \{-1,1\}$, matrix $\bbA$ and $\bbD_d$, intermediate variables $\bbz_1^k,\bbz_2^k,\bbu_1^k,\bbu_2^k$, and hyperparameter $\lambda_3$
\State \textbf{Initialize}: $\bbx \gets \bbx^k$, $\rho > 0$, 
and stopping threshold $\text{TOLERANCE} = 10^{-5}$.
\Repeat
    \State
    $\nabla \bbL_\bbx \gets -\sum_{i=1}^N \frac{y_i \bba_i}{1 + \exp(y_i \bba_i^\top \bbx)} + \rho ( (1 +\frac{\lambda_3}{\rho}) \bbx - \bbz_1^k + \frac{\bbu_1^k}{\rho}) + \rho \bbD_d^\top (\bbD_d \bbx - \bbz_2^k + \frac{\bbu_2^k}{\rho})$  \Comment{Gradient}
    \State $\bbH_\bbx \gets \sum_{i=1}^N \frac{\exp(y_i \bba_i^\top \bbx)}{(1 + \exp(y_i \bba_i^\top \bbx))^2} \bba_i \bba_i^\top + (\rho + \lambda_3) \bbI + \rho \bbD_d^\top \bbD_d$  \label{eq:alg_hessian}\Comment{Hessian}
    \State $ \Delta \bbx \gets -\bbH_\bbx^{-1} \nabla \bbL_\bbx $
    \Comment{Newton direction}
    \State $ \eta \gets -(\nabla \bbL_\bbx)^\top \Delta \bbx $  \Comment{Newton decrement}
    \State \textbf{If} $ |\eta| \leq \text{TOLERANCE}$, \textbf{then} break
    \State Find step size $t$ using backtracking line search~\cite{boyd2004ConvexOptimizationBook}
    \State $ \bbx \gets  \left[ \bbx + t \Delta \bbx \right]_+$
\Until convergence
\end{algorithmic}
\label{alg:newtonforadmm}
\end{algorithm}

\noindent  \underline{2. $\bbz_1$-update ($\ell_1$ projection)}
\begin{align}
    \bbz_1^{k+1} & = \argmin_{\bbz_1}L(\bbx^{k+1}, \bbz_1, \bbz_2^k, \bbu_1^k, \bbu_2^k) \notag \\
    & = \argmin_{\bbz_1} \lambda_1 \|\bbz_1\|_1 + \frac\rho 2 \left\|\bbz_1 - \big(\bbx^{k+1} + \frac1\rho \bbu_1^k\big)\right\|_2^2.
    \label{eq:admm_auxiliary_z1}
\end{align}
The solution to~\eqref{eq:admm_auxiliary_z1} is the computation of the corresponding proximal operator, that is
\begin{align}
    \bbz_1^{k+1} = \prox_{\frac{\lambda_1}{\rho} \|\cdot\|_1} 
    \left( \bbx^{k+1} + \frac1\rho \bbu_1^{k}\right)
\end{align}
where the proximal operator for the $\ell_1$ norm is a soft-thresholding function
\begin{align}
    \prox_{\beta \|\cdot\|_1}(\bbv) 
    = \operatorname{soft}_{\beta \|\cdot\|_1}(\bbv)
    = \sign(\bbv)\odot\max(|\bbv| - \beta, 0).
\end{align}
This operation promotes sparsity in the solution by shrinking small-magnitude components toward zero.

\noindent \underline{3. $\bbz_2$-update ($\ell_1$ projection)}
\begin{align}
    \bbz_2^{k+1} & = \argmin_{\bbz_2}L(\bbx^{k+1}, \bbz_1^{k+1}, \bbz_2, \bbu_1^k, \bbu_2^k) \notag \\
    & =  \argmin_{\bbz_2} \lambda_2 \|\bbz_2\|_1 + \frac\rho2 \|\bbz_2 - (\bbD\bbx^{k+1}+\frac1\rho \bbu_2^k)\|_2^2.
\end{align}
The solution is 
\begin{align}
    \bbz_2^{k+1} = \prox_{\frac{\lambda_2}{\rho} \|\cdot\|_1} 
    \left( \bbD\bbx^{k+1} + \frac1\rho \bbu_2^{k} \right),
\end{align}
which uses the same soft-thresholding operator.
This update enforces piecewise constancy by encouraging sparsity in the differences between adjacent elements.

\noindent \underline{4. $\bbu$-updates (Dual Update)}
\begin{align}
    \bbu_1^{k+1} & = \bbu_1^k + \rho(\bbx^{k+1} - \bbz_1^{k+1}), \\
    \bbu_2^{k+1} & = \bbu_2^k + \rho(\bbD\bbx^{k+1} - \bbz_2^{k+1}).
\end{align}
These updates adjust the dual variables to drive the solution toward primal feasibility.

\noindent \underline{5. Stopping Criteria}\\
We monitor both primal and dual residuals to determine convergence
\begin{align}
    \bbr_1^k & = \bbx^{k+1} - \bbz_1^{k+1}, \bbr_2^k = \bbD\bbx^{k+1} - \bbz_2^{k+1}, \\
    \bbs_1^k & = \rho (\bbx^{k+1} - \bbz_1^{k+1}), \bbs_2^k = \rho (\bbD\bbx^{k+1} - \bbz_2^{k+1}).
\end{align}
The algorithm terminates when
\begin{align}
    \max\left(\|\bbr_1^k\|_2, \|\bbr_2^k\|_2, \|\bbs_1^k\|_2, \|\bbs_2^k\|_2 \right) \leq \epsilon,
\end{align}
where $\epsilon >0$ is a predefined tolerance threshold.
The complete ADMM procedure is summarized in Algorithm~\ref{alg:admm_nnfl}.

\begin{algorithm}[!t]
	\caption{ADMM for subproblem}
	\begin{algorithmic}[1]
        \State \textbf{Input:} Response $\bby$, matrix $\bbA$ and $\bbD_d$, hyper-parameters $(\lambda_1, \lambda_2, \lambda_3)$.
        \State \textbf{Initialize:} Randomize $(\bbx^0, \bbz_1^0, \bbz_2^0, \bbu_1^0, \bbu_2^0)$, $\rho > 0$, $\epsilon = 10^{-5}$
        \For{$k=0,1,2, \ldots$}
            \If{$\mathsf{Linear \ regression}$}
            \State $ \bbM \gets \bbA^\top\bbA + (\rho + \lambda_3) \bbI + \rho \bbD_d^\top\bbD_d$ \Comment{Compute only once}
            \State $\bbb^k \gets \bbA^\top\bby + \rho (\bbz_1^k - \bbu^k_1/\rho) + \rho \bbD_d^\top (\bbz^k_2 - \bbu^k_2/\rho)$
            \State $ \bbx^{k+1} \gets \left[ \bbM^\inv \bbb^k \right]_+ $
            \ElsIf{$\mathsf{Logistic\ regression}$}
            \State Solve $\bbx^{k+1}$ using Newton's method using Algorithm~\ref{alg:newtonforadmm}
            \EndIf
            \State $ \bbz_1^{k+1} \gets \prox_{\frac{\lambda_1}{\rho} \|\cdot\|_1}  \left( \bbx^{k+1} + \frac1\rho \bbu_1^{k} \right) $
            \State $ \bbz_2^{k+1} \gets \prox_{\frac{\lambda_2}{\rho} \|\cdot\|_1} \left( \bbD_d\bbx^{k+1} + \frac1\rho \bbu_2^{k} \right) $
            \State $ \bbu_1^{k+1} \gets \bbu_1^k + \rho (\bbx^{k+1} - \bbz_1^{k+1}) $
            \State $ \bbu_2^{k+1} \gets \bbu_2^k + \rho (\bbD_d\bbx^{k+1} - \bbz_2^{k+1}) $
            \State $\bbr_1^{k+1} \gets \bbx^{k+1} - \bbz_1^{k+1}$, $\bbr_2^{k+1} \gets \bbD_d\bbx^{k+1} - \bbz_2^{k+1}$, $\bbs_1^{k+1} \gets \rho (\bbz_1^{k+1} - \bbz_1^{k})$,  $\bbs_2^{k+1} \gets \rho (\bbz_2^{k+1} - \bbz_2^{k})$
            \State \textbf{If} 
            $\max\left(\|\bbr_1^{k+1}\|_2, \|\bbr_2^{k+1}\|_2, \|\bbs_1^{k+1}\|_2, \|\bbs_2^{k+1}\|_2 \right) \leq \epsilon$ \textbf{then} break
        \EndFor
	\end{algorithmic}
	\label{alg:admm_nnfl}
\end{algorithm}

The convergence of the proposed NS-KTR algorithm can be established based on the general theory for alternating optimization methods applied to multilinear optimization problems, as discussed in~\cite{huang2016flexible}.
In our framework, each subproblem in the alternating optimization procedure is convex with respect to the individual factor vectors when others are fixed, and the inclusion of a ridge regularization term ($\lambda_{d3} > 0$) renders them strongly convex, ensuring unique solutions.
Consequently, following the convergence results for block coordinate descent methods~\cite{Hong2017ADMMConvergence}, every limit point of the sequence generated by the alternating optimization algorithm is a stationary point of the overall tensor regression objective.
Additionally, the ADMM algorithm (Algorithm~\ref{alg:admm_nnfl}) used to solve each subproblem converges to the optimal solution of that subproblem, given its convexity~\cite{Eckstein1992Apr}. 
As explained above the sequence of iterates remains bounded, thus, the NS-KTR algorithm converges to a stationary point.

\subsection{Rank Selection}
\label{secsub:rankBIC}

An essential aspect of tensor regression models is determining the appropriate rank $R$ for the CP decomposition. 
The rank parameter controls the model complexity and directly impacts both computational efficiency and predictive performance. 
While increasing the rank allows us to capture more complex patterns in the data, it also risks overfitting and unnecessarily increases the number of parameters.

Following~\cite{zhou13tensor}, we employ the Bayesian Information Criterion (BIC)~\cite{schwarz1978BIC} as our model selection method, which provides a principled approach to balancing model fit and complexity. 
The BIC is defined as
\begin{align}
    \label{eq:BIC}
    \BIC = -2 L(\hat\bbtheta) + \log(N)p_e,
\end{align}
where $L(\hat\bbtheta)$ represents the maximized log-likelihood function, parameter $\hat \bbtheta$ consists of optimal values $\{ \hat \bbB_d\}_{d=1}^D$ obtained by running the NS-KTR algorithm, and $p_e = R(\sum_{d=1}^D I_d - D + 1)$ denotes the effective number of parameters in the model. 
This formulation for $p_e$ accounts for the total parameters in the CPD while considering the scaling indeterminacy, i.e., each factor matrix $\bbB_d \in \mathbb{R}^{I_d \times R}$ contributes $I_dR$ parameters, but we subtract $R(D-1)$ constraints due to the scaling ambiguity inherent in the CPD.

For linear regression, the log-likelihood is defined as 
\begin{align}
    L(\hbtheta) = -\frac{N}{2}\log(2\pi\hat{\sigma}^2) - \frac N2,
\end{align}
where $\hat{\sigma}^2 = \frac1N \sum_{i=1}^N(y_i - \langle\mathcal{X}_i, \hat{\mathcal{B}}\rangle)^2$ is the estimated error variance.

For logistic regression, when using encoding $\{-1,1\}$ for the targets, the log-likelihood is defined as
\begin{align}
    L(\hbtheta) = \sum_{i=1}^N \left[y_i\log(p_i) + (1-y_i)\log(1-p_i)\right],
\end{align}
where $p_i = \frac{1}{1+\exp(-\langle\mathcal{X}_i, \hat{\mathcal{B}}\rangle)}$ is the predicted probability for sample $i$.

The BIC effectively penalizes complex models, particularly as the sample size increases. 
A lower BIC value indicates a better balance between model fit and complexity.
In practice, we fit multiple models with different rank values and select the rank that minimizes the BIC. 

\section{Numerical Experiments}
\label{sec:num_exp}
In this section, we evaluate the empirical performance of the proposed NS-KTR on both synthetic and real HSI datasets. 
The synthetic experiments quantify estimation accuracy under controlled conditions with varying noise levels, sample sizes, and rank selections. 
We employ both 2D and 3D tensor signals with diverse structural characteristics to systematically assess the model's performance across different signal patterns. 
These experiments enable direct comparison of various regularization approaches within the NS-KTR framework.
For real-data validation, we apply NS-KTR to HSI datasets of crop leaves, demonstrating the method's efficacy in practical applications where signal structures exhibit complex spatial-spectral patterns. 
These experiments evaluate both linear and logistic regression variants of the proposed model, for regression and classification tasks respectively.

\subsection{Synthetic Dataset}
\begin{figure}[!t]
    \centering
    \centerline{\includegraphics[width = 0.99\linewidth]{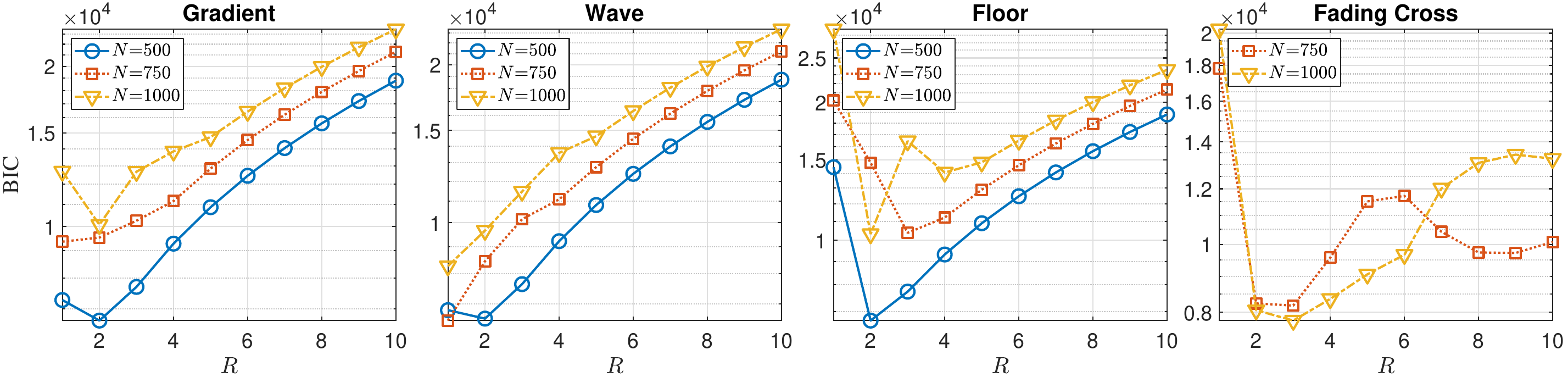}}
    \caption{Rank selection through BIC for responses generated by different signal parameters~$\ccalB$ with number of responses $N$ varying in $[500, 750, 1000]$.}
    \label{fig:BIC_rank}
\end{figure}

\begin{figure*}[!h]
    \centering
    \subfloat{\includegraphics[width = 0.95\linewidth]{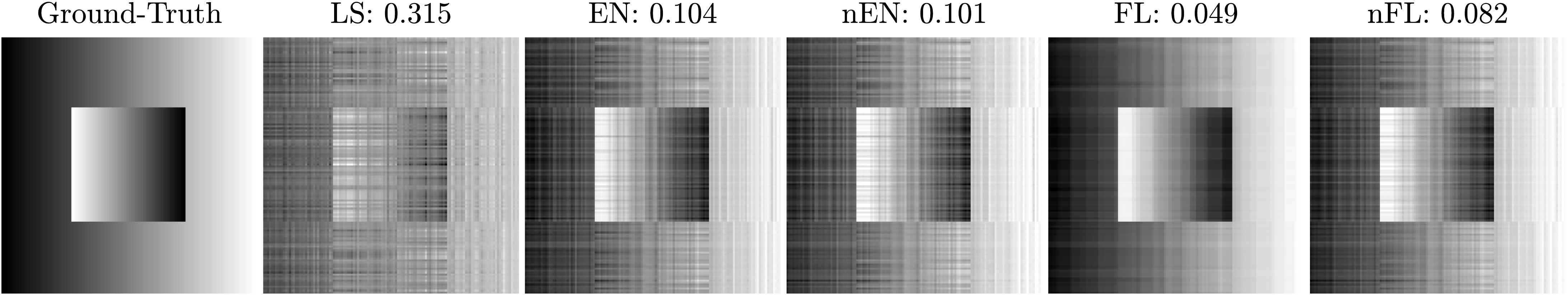}}\\
    %
    \subfloat{\includegraphics[width = 0.95\linewidth]{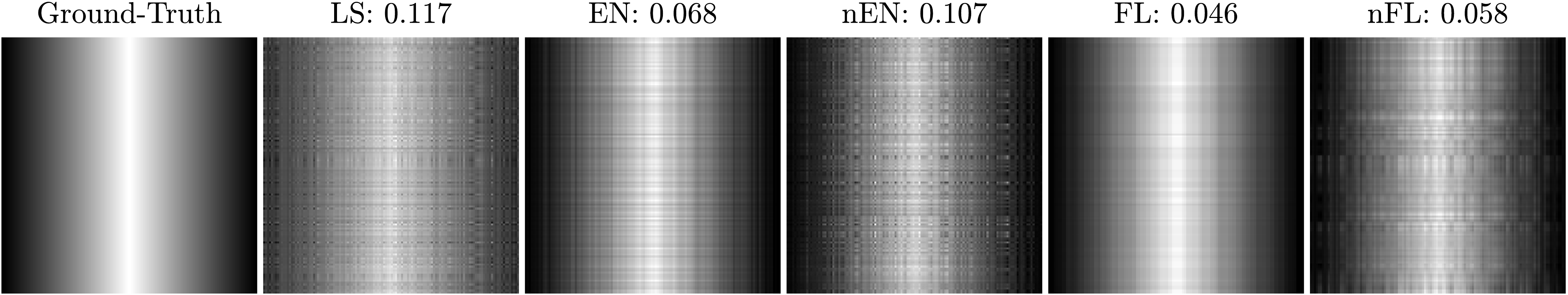}}\\
    \subfloat{\includegraphics[width = 0.95\linewidth]{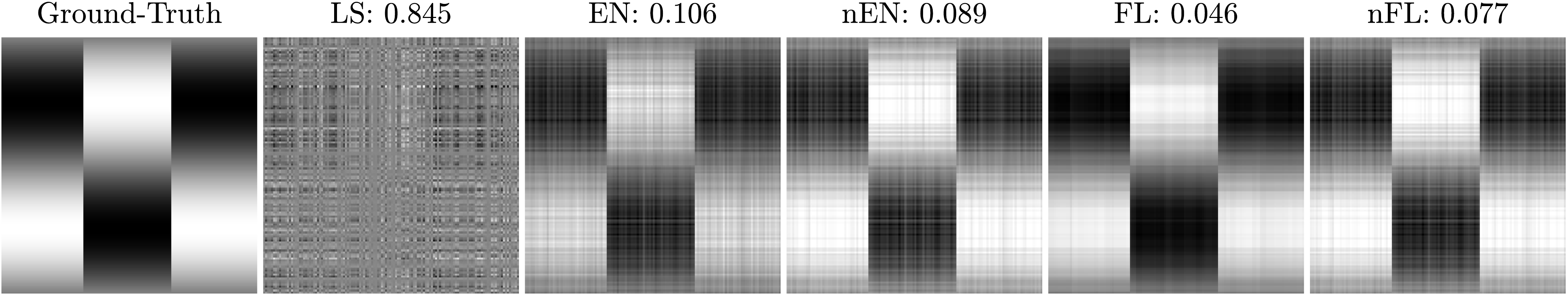}}\\
    \subfloat{\includegraphics[width = 0.95\linewidth]{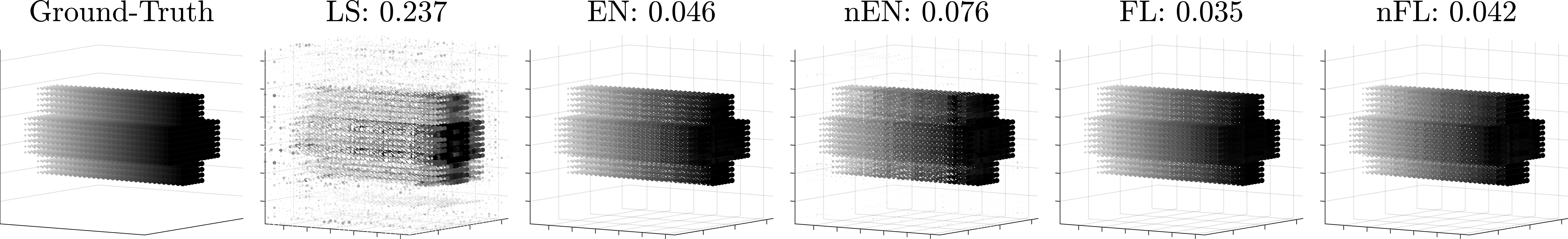}}
    \caption{Visual comparison of tensor signal reconstruction using different regularization approaches within the NSKTR framework. 
    The first column shows the true signals (Gradient, Floor, Wave, and Fading Cross), with subsequent columns displaying reconstructions using LS, EN, nEN, FL, and nFL regularization. 
    Results represent median estimation error with $N=1000$ samples, SNR=$20$dB, $R=2$ for 2D signals, and $R=3$ for the 3D signal.}
    \label{fig:2Dand3Dcomp}
\end{figure*}

{\it Setup.} 
The response variables are generated using the model $y_i = \langle \mathcal{X}_i, \mathcal{B} \rangle + e_i$, where the tensor covariates $\mathcal{X}_i$ are drawn from a standard normal distribution, and $e_i, i=1,\ldots,N$ are i.i.d. Gaussian noise terms. 
Two noise levels are tested at Signal-to-Noise Ratios (SNR) of 20 dB and 10 dB to evaluate performance under different noise conditions.
We define SNR as 
$
    \textrm{SNR} = 10 \log\left(\frac{\sum_{i=1}^N \langle \mathcal{X}_i, \mathcal{B} \rangle^2}{\sum_{i=1}^N e_i^2} \right).
$
The true tensor parameter $\mathcal{B}$ comprises four synthetic signals: Gradient, Floor, and Wave for the 2D case ($I_1 = I_2 =128$), and Fading Cross for the 3D case ($I_1 = I_2 = I_3 = 32$).
These signals exhibit varying degrees of smoothness, sparsity, and structural complexity to test the method's efficacy across diverse signal characteristics. 
The first column of Fig.~\ref{fig:2Dand3Dcomp} displays the 2D and 3D tensor signals $\mathcal{B}$.

{\it Estimation criterion.} 
The performance is evaluated using the estimation error (EE), defined as $\text{EE} = {\|\hat{\mathcal{B}} - \mathcal{B}\|_\fronorm}/{\|\mathcal{B}\|_\fronorm}$, where $\hat{\mathcal{B}}$ denotes the estimated tensor signal.
The proposed NS-KTR framework is implemented with five different regularization configurations: Least Squares (LS), Elastic Net (EN), Nonnegative Elastic Net (nEN), Fused LASSO (FL), and Nonnegative Fused LASSO (nFL).
These configurations represent varying combinations of regularization within our framework, from unstructured (LS, EN) to structured (FL) and nonnegative-constrained (nEN, nFL) approaches.
The LS configuration corresponds to the KTR method in~\cite{zhou13tensor}, implemented by setting all penalty parameters to zero for consistent comparison.

{\it Rank selection.} 
Sample sizes vary as $N = [500, 750, 1000]$ for 2D signals and $N = [750, 1000]$ for the 3D signal.
For each signal type, we select the rank corresponding to the minimum BIC value, as described in Section~\ref{secsub:rankBIC}, along with two adjacent ranks—typically one rank lower and one rank higher than the optimal choice.
Fig.~\ref{fig:BIC_rank} shows the BIC curves for the four tensor signals, with each curve representing mean BIC values from $50$ Monte Carlo simulations.

{\it Regularization parameter selection.} 
The performance of structured tensor regression models is sensitive to the choice of regularization parameters $\{\lambda_{d1}, \lambda_{d2}, \lambda_{d3}\}_{d=1}^D$. 
A grid $\{\lambda^{(i)}\}_{i=0}^{L+1}$ is defined with equispaced log-scale intervals: $\lambda^{(L+1)} = 0$, $\lambda^{(L)} = \epsilon \lambda^{(0)}$, $\lambda^{(j)} = \epsilon^{j/L} \lambda^{(0)} = \epsilon^{1/L} \lambda^{(j-1)}$, yielding $\log(\lambda^{(j-1)}) - \log(\lambda^{(j)}) = \frac{\log(\lambda^{(0)}) - \log(\lambda^{(L)})}{L}$. 
We set $\lambda^{(0)} = N$, $\epsilon = 10^{-3}$, and $L = 5$ for all experiments. 
A sequential greedy grid search optimizes these parameters: starting with $d=1$, $\lambda_{d1}$ (sparsity) is tuned while fixing $\lambda_{d2} = \lambda_{d3} = 0$; then, fixing $\lambda_{d1}$ and $\lambda_{d3}$, $\lambda_{d2}$ (piecewise constancy) is optimized; finally, $\lambda_{d3}$ (smoothness) is tuned. 
This approach scales linearly with the number of dimension $D$, reducing computational cost compared to a full grid search. 
The tuned parameters reflect signal characteristics, e.g., high $\lambda_{d3}$ for smoothness (Gradient) or high $\lambda_{d2}$ for piecewise constancy (Floor)—offering adaptability over fixed-prior methods (e.g.,~\cite{zhou13tensor,zhou2014regmtx}) and improving estimation accuracy.

\begin{table}[!h]
\centering
\resizebox{0.999\linewidth}{!}{%
\begin{tabular}{cccccccc|cccccc}
\hline \hline
\multirow{2}{*}{Data Name} & \multirow{2}{*}{$n$} & \multirow{2}{*}{Rank} & \multicolumn{5}{c}{SNR = 20dB} & \multicolumn{5}{c}{SNR = 10dB} \\
\cmidrule(lr){4-8} \cmidrule(lr){9-13}
& & & LS & EN & nEN & FL & nFL & LS & EN & nEN & FL & nFL \\
\midrule
\multirow{9}{*}{{\it Gradient}}
& \multirow{3}{*}{500}
& 1 & 1.25(0.49) & 1.03(0.59) & 0.35(0.01) & 0.88(0.61) & \blue{0.32(0.01)} 
      & 1.46(0.30) & 1.25(0.50) & 0.48(0.03) & 1.11(0.56) & \blue{0.43(0.02)} \\
& & 2 & 2.18(0.12) & 1.56(0.12) & 0.51(0.03) & 1.25(0.44) & \blue{0.16(0.04)} 
      & 2.24(0.11) & 1.64(0.10) & 0.67(0.04) & 1.49(0.23) & \blue{0.50(0.06)} \\
& & 3 & 2.56(0.20) & 1.12(0.15) & 0.51(0.03) & 0.33(0.39) & \blue{0.15(0.02)} 
      & 2.62(0.20) & 1.23(0.11) & 0.99(0.05) & \blue{0.79(0.34)} & 0.95(0.07) \\
\cmidrule(lr){2-13}
\multirow{3}{*}{} 
& \multirow{3}{*}{750}
& 1 & 0.42(0.30) & 0.32(0.15) & 0.30(0.01) & 0.29(0.14) & \blue{0.29(0.00)}
      & 0.53(0.32) & 0.39(0.15) & 0.36(0.01) & \blue{0.34(0.06)} & 0.35(0.01) \\
& & 2 & 1.28(0.50) & 0.68(0.55) & 0.26(0.09) & 0.25(0.46) & \blue{0.14(0.06)} 
      & 1.50(0.39) & 0.89(0.45) & 0.62(0.03) & 0.53(0.43) & \blue{0.48(0.07)} \\
& & 3 & 2.08(0.14) & 0.86(0.41) & 0.31(0.05) & \blue{0.11(0.19)} & 0.13(0.03) 
      & 2.19(0.17) & 1.17(0.32) & 0.66(0.04) & 0.49(0.38) & \blue{0.40(0.04)} \\
\cmidrule(lr){2-13}
\multirow{3}{*}{} 
& \multirow{3}{*}{1000}
& 1 & 0.35(0.08) & 0.33(0.01) & 0.33(0.01) & \blue{0.30(0.01)} & 0.32(0.01) 
      & 0.35(0.08) & 0.33(0.01) & 0.33(0.01) & \blue{0.30(0.01)} & 0.32(0.01) \\
& & 2 & 0.33(0.26) & 0.11(0.03) & 0.11(0.04) & \blue{0.05(0.00)} & 0.08(0.00) 
      & 0.55(0.14) & 0.42(0.05) & 0.39(0.06) & \blue{0.21(0.03)} & 0.31(0.04) \\
& & 3 & 1.06(0.51) & 0.30(0.39) & 0.15(0.04) & \blue{0.06(0.01)} & 0.10(0.01) 
      & 1.39(0.40) & 0.60(0.28) & 0.48(0.04) & \blue{0.25(0.21)} & 0.34(0.03) \\
\midrule
\multirow{9}{*}{{\it Floor}}
& \multirow{3}{*}{500}
& 1 & 1.14(0.57) & 0.72(0.67) & 0.10(0.01) & 0.54(0.68) & \blue{0.08(0.01)} 
      & 1.34(0.40) & 0.90(0.57) & 0.32(0.02) & 0.68(0.60) & \blue{0.27(0.02)} \\
& & 2 & 2.15(0.13) & 1.44(0.25) & 0.25(0.02) & 0.93(0.65) & \blue{0.10(0.01)} 
      & 2.27(0.15) & 1.61(0.13) & 0.62(0.04) & 1.34(0.40) & \blue{0.37(0.02)} \\
& & 3 & 2.53(0.18) & 1.02(0.19) & 0.40(0.04) & \blue{0.07(0.08)} & 0.14(0.01) 
      & 2.69(0.19) & 1.14(0.13) & 0.55(0.03) & 0.44(0.32) & \blue{0.42(0.03)} \\
\cmidrule(lr){2-13}
\multirow{3}{*}{}
& \multirow{3}{*}{750}
& 1 & 0.21(0.38) & 0.10(0.20) & 0.07(0.00) & 0.07(0.19) & \blue{0.06(0.00)} 
      & 0.41(0.40) & 0.27(0.21) & 0.22(0.01) & \blue{0.20(0.15)} & 0.21(0.01) \\
& & 2 & 1.14(0.67) & 0.49(0.64) & 0.12(0.02) & 0.20(0.45) & \blue{0.08(0.01)} 
      & 1.38(0.46) & 0.60(0.42) & 0.38(0.02) & \blue{0.29(0.27)} & 0.30(0.04) \\
& & 3 & 2.10(0.17) & 0.44(0.48) & 0.15(0.01) & 0.07(0.17) & \blue{0.07(0.01)} 
      & 2.19(0.15) & 0.97(0.41) & 0.43(0.02) & 0.32(0.32) & \blue{0.32(0.02)} \\
\cmidrule(lr){2-13}
\multirow{3}{*}{}
& \multirow{3}{*}{1000}
& 1 & 0.06(0.03) & 0.06(0.00) & 0.06(0.00) & \blue{0.04(0.00)} & 0.49(0.48) 
      & 0.20(0.10) & 0.19(0.01) & 0.18(0.01) & \blue{0.14(0.01)} & 0.17(0.01) \\
& & 2 & 0.21(0.35) & 0.10(0.22) & 0.13(0.18) & \blue{0.05(0.00)} & 0.07(0.01) 
      & 0.42(0.24) & 0.31(0.02) & 0.30(0.02) & \blue{0.20(0.02)} & 0.21(0.05) \\
& & 3 & 0.82(0.64) & 0.18(0.38) & 0.10(0.01) & \blue{0.04(0.00)} & 0.06(0.01) 
      & 1.22(0.44) & 0.34(0.04) & 0.36(0.02) & \blue{0.16(0.02)} & 0.29(0.02) \\
\midrule
\multirow{9}{*}{{\it Wave}}
& \multirow{3}{*}{500}
& 1 & 1.51(0.17) & 1.41(0.24) & 0.80(0.04) & 1.33(0.31) & \blue{0.75(0.04)} 
      & 1.57(0.13) & 1.50(0.17) & 0.92(0.04) & 1.44(0.22) & \blue{0.85(0.04)} \\
& & 2 & 2.20(0.13) & 1.62(0.10) & 0.93(0.09) & 1.52(0.16) & \blue{0.73(0.29)} 
      & 2.26(0.14) & 1.69(0.10) & 1.02(0.05) & 1.62(0.13) & \blue{1.04(0.07)} \\
& & 3 & 2.57(0.21) & 1.25(0.07) & 1.04(0.06) & 0.86(0.36) & \blue{0.18(0.14)} 
      & 2.64(0.18) & 1.31(0.06) & 0.90(0.03) & 1.13(0.22) & \blue{0.53(0.12)} \\
\cmidrule(lr){2-13}
\multirow{3}{*}{}
& \multirow{3}{*}{750}
& 1 & 0.93(0.26) & 0.71(0.14) & 0.66(0.02) & 0.67(0.14) & \blue{0.65(0.02)} 
      & 0.99(0.27) & 0.80(0.20) & 0.71(0.02) & 0.74(0.18) & \blue{0.69(0.02)} \\
& & 2 & 1.65(0.22) & 1.29(0.37) & 0.35(0.30) & 0.67(0.62) & \blue{0.27(0.28)} 
      & 1.70(0.19) & 1.39(0.28) & 0.55(0.22) & 0.93(0.51) & \blue{0.37(0.16)} \\
& & 3 & 2.13(0.12) & 1.32(0.21) & 0.34(0.18) & 0.50(0.56) & \blue{0.41(0.26)} 
      & 2.22(0.14) & 1.47(0.13) & 0.71(0.13) & 1.00(0.35) & \blue{0.38(0.04)} \\
\cmidrule(lr){2-13}
\multirow{3}{*}{}
& \multirow{3}{*}{1000}
& 1 & 0.63(0.05) & 0.62(0.01) & 0.61(0.01) & \blue{0.60(0.01)} & 0.61(0.01) 
      & 0.70(0.14) & 0.65(0.01) & 0.64(0.01) & \blue{0.63(0.01)} & 0.64(0.01) \\
& & 2 & 0.80(0.43) & 0.26(0.37) & 0.16(0.19) & \blue{0.05(0.00)} & 0.15(0.19) 
      & 0.94(0.33) & 0.41(0.16) & 0.28(0.07) & \blue{0.21(0.02)} & 0.28(0.12) \\
& & 3 & 1.55(0.30) & 0.56(0.49) & 0.15(0.09) & \blue{0.09(0.18)} & 0.11(0.09) 
      & 1.77(0.21) & 1.16(0.33) & 0.43(0.12) & 0.34(0.28) & \blue{0.33(0.02)} \\
\midrule
\multirow{6}{*}{{\it Fading Cross}}
& \multirow{3}{*}{750}
& 2 & 0.72(0.47) & 0.46(0.49) & 0.24(0.23) & 0.37(0.45) & \blue{0.20(0.18)} 
      & 0.93(0.42) & 0.68(0.42) & 0.19(0.08) & 0.49(0.46) & \blue{0.18(0.06)} \\
& & 3 & 0.93(0.50) & 0.56(0.59) & 0.05(0.01) & 0.35(0.53) & \blue{0.07(0.09)}
      & 0.98(0.45) & 0.71(0.54) & 0.21(0.07) & 0.59(0.56) & \blue{0.20(0.04)} \\
& & 4 & 1.09(0.54) & 0.60(0.64) & \blue{0.06(0.02)} & 0.17(0.37) & 0.08(0.09) 
      & 1.34(0.38) & 1.01(0.60) & 0.26(0.05) & 0.71(0.62) & \blue{0.21(0.02)} \\
\cmidrule(lr){2-13}
\multirow{3}{*}{} 
& \multirow{3}{*}{1000}
& 2 & 0.44(0.41) & 0.28(0.44) & \blue{0.04(0.00)} & 0.21(0.37) & \blue{0.04(0.00)} 
      & 0.56(0.35) & 0.33(0.29) & 0.24(0.14) & \blue{0.23(0.29)} & 0.25(0.15) \\
& & 3 & 0.30(0.38) & 0.04(0.00) & 0.09(0.11) & \blue{0.03(0.00)} & 0.07(0.09) 
      & 0.63(0.40) & 0.30(0.31) & 0.18(0.08) & \blue{0.13(0.09)} & 0.20(0.12) \\
& & 4 & 0.56(0.57) & 0.26(0.49) & 0.08(0.01) & 0.09(0.25) & \blue{0.07(0.09)} 
      & 0.85(0.44) & 0.32(0.31) & 0.19(0.02) & \blue{0.15(0.24)} & 0.18(0.03) \\
\hline \hline
\end{tabular}
}
\caption{Estimation error (EE) comparison for different regularization approaches within the NSKTR framework. Results report mean(standard deviation) for various signal types, sample sizes ($N$), ranks, and SNR levels.
The method yielding the lowest mean of EE for each configuration is highlighted in \blue{blue}.}
\label{tab:syntheticEE}
\end{table}

Figure~\ref{fig:2Dand3Dcomp} presents the visual comparison of signal parameter reconstructions using different regularization approaches within our proposed NS-KTR framework. 
The first column displays the true signals (Gradient, Floor, Wave, and Fading Cross), with subsequent columns showing reconstructions using LS, EN, nEN, FL, and nFL regularization.
The visual comparisons reveal how different regularization strategies capture the underlying signal structures. 
For the Gradient signal, LS and EN introduce more noise artifacts, while regularized methods better preserve continuous intensity transitions. 
In the Floor signal, the visualizations show varying capabilities in edge preservation and region homogeneity, with regularization approaches improving boundary definition. 
The Wave signal demonstrates how methods differ in capturing periodic patterns—some better preserving fine details while others emphasizing pattern continuity. 
For the 3D Fading Cross, methods with nonnegativity constraints show distinct characteristics in reconstructing the volumetric structure and intensity gradation throughout the shape.

Table~\ref{tab:syntheticEE} quantitatively supports these visual observations, reporting mean estimation errors with standard deviations across varying sample sizes, ranks, and SNR levels. 
Methods with nonnegativity constraints generally perform better than their unconstrained counterparts in most settings, though the optimal method varies by signal type and experimental conditions. 
Notably, even in cases where nonnegative methods don't achieve the lowest mean error, they consistently exhibit smaller standard deviations, indicating more stable performance across 50 Monte Carlo simulations.
For example, with the Floor signal at $N=1000$ and $R=3$, nFL achieves an error of 0.06(0.01) at 20dB SNR compared to 0.18(0.38) for EN, where the substantially smaller standard deviation (0.01 vs. 0.38) demonstrates superior consistency of the nonnegative approach.

The advantage of structured regularization becomes more evident as sample size increases from 500 to 1000. This is observed in the Wave signal, where nFL's error decreases from 0.75(0.04) to 0.61(0.01) for rank 1 at 20dB SNR. Lower ranks typically yield better results with limited samples, consistent with our BIC analysis in Figure~\ref{fig:BIC_rank}. 
Additionally, under higher noise conditions (10dB vs. 20dB), regularized methods demonstrate better robustness compared to LS, with nonnegative constraints providing more consistent results as evidenced by their smaller error variances.
The combined analysis of Figure~\ref{fig:2Dand3Dcomp} and Table~\ref{tab:syntheticEE} shows that the algorithm performance depends on signal characteristics. 
For Gradient signals, both FL-family and nEN methods perform well depending on sample size. 
Floor signals with piecewise-constant regions often benefit from FL-based regularization at higher sample sizes. 
For Wave signals, performance varies with sample size and rank. 
The 3D Fading Cross results indicate that nonnegativity constraints significantly improve reconstruction of volumetric data.
These results demonstrate that our NS-KTR framework provides adaptability across various signal types and experimental conditions while delivering greater estimation stability.

\subsection{Real HSI Data Applications}
\begin{figure}[!t]
    \centering
    \centerline{\includegraphics[width = 0.99\linewidth]{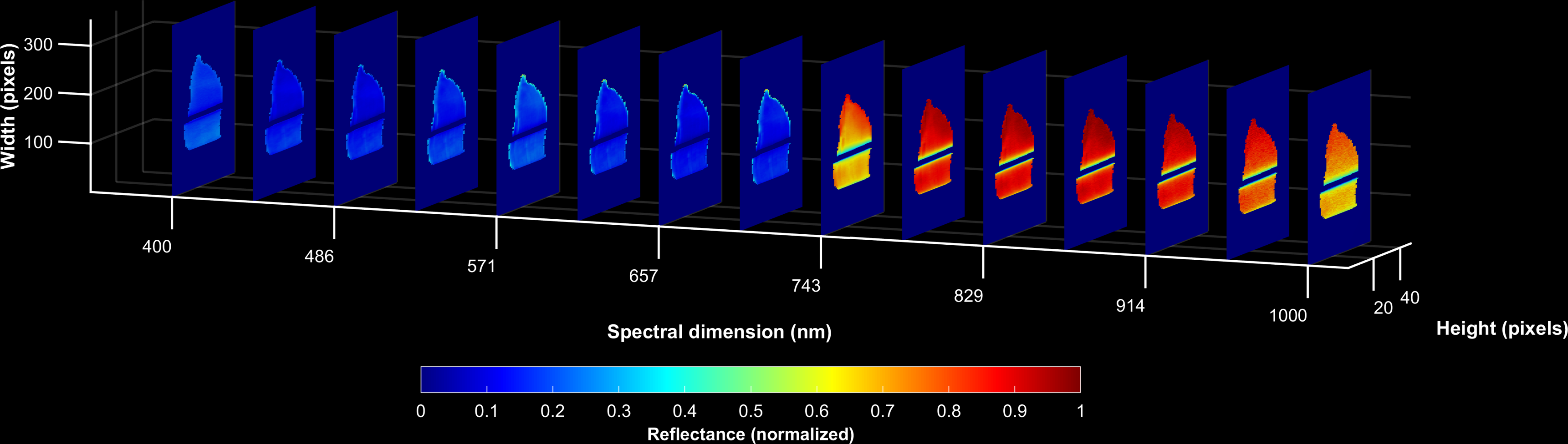}}
    \caption{3D visualization of the $199$th image from the HyperLeaf2024 dataset. 
    The visualization displays $15$ spectral bands (selected uniformly from 400nm to 1000nm) as parallel slices along the x-axis. 
    Each slice represents a 2D spatial image (48×352 pixels) at a specific wavelength. 
    The y-axis represents the height dimension, the z-axis represents the width dimension, and the x-axis represents the spectral dimension.
    Color intensity corresponds to normalized reflectance values, with brighter regions indicating higher reflectance.}
    \label{fig:Hyper}
\end{figure}

HSI whole-image regression presents technical challenges compared to pixel-wise approaches. 
Whole-image analysis requires modeling complex spatial-spectral interactions across the entire domain while effectively managing high-dimensional data structures. 
Unlike pixel-wise methods where each spatial location is analyzed independently, whole-image analysis must capture the intricate relationships between spectral signatures and their spatial distribution throughout the image.

{\it Dataset.} 
We evaluate our methods on HyperLeaf2024 dataset~\cite{laprade2024hyperleaf2024} of wheat flag leaves.
The dataset contains 2410 hyperspectral images, with each image represented as a tensor with dimensions $128\times128\times224$, where the third dimension corresponds to 224 spectral bands. 
Each hyperspectral image captures spatial-spectral information about individual wheat leaves, with images taken from various plots with different growing conditions.
A 3D visualization of the $199$th image of this dataset is give in Fig.~\ref{fig:Hyper}.
Since the official test set is reserved for Kaggle competition evaluation with limited access, we divided the original 1590 training data using a 6:2:2 ratio for training, validation, and testing, respectively. 
This strategy allows us to perform parameter tuning on the validation set and evaluate final model performance on our test set. 

The dataset includes eight target variables, which we categorize into regression and classification tasks: \\
\indent -- Targets 1-4 for regression: GrainWeight (grain yield measurement), Gsw (stomatal conductance related to plant water use efficiency), \ \ PhiPS2 (chlorophyll fluorescence indicating photosynthetic efficiency), and Fertilizer (content with discrete levels 0.0, 0.5, 1.0). \\
\indent -- Targets 5-8 for classification: wheat cultivars (Heerup, Kvium, Rembrandt, Sheriff) encoded as one-hot variables.
We apply the linear regression model on Targets 1-4 and the logistic regression model on Targets 5-8.

We compare our NS-KTR method with three established tensor regression approaches: \textit{(i)} Tensor linear regression (TLR)~\cite{zhou13tensor}, which provides a basic tensor regression framework without specialized regularization; \textit{(ii)} Tucker tensor regression (TuckerTR)~\cite{li2018tucker_comp2}, which uses Tucker decomposition with a more flexible core tensor; and \textit{(iii)} Tensor quantile regression (TQtR)~\cite{li2021tensorquantile}, which extends quantile regression to the tensor setting for robust estimation.
\revised{
These three models represent the principal paradigms of tensor regression, Kruskal-based, Tucker-based, and quantile-based formulations, and thus, provide representative benchmarks for evaluating the proposed NS-KTR.
}

HSI data possesses distinct structural characteristics across different dimensions that make our proposed mode-wise regularization particularly effective. 
The spectral dimension (mode 3) typically exhibits smooth variations across adjacent wavelengths, with certain wavelength regions demonstrating strong correlations. This aligns with our ridge regularization ($\lambda_3$), which promotes smoothness. 
Spatial dimensions (modes 1-2) often contain locally constant profiles with meaningful boundaries between different tissue types, matching the characteristics targeted by total variation regularization ($\lambda_2$). Additionally, only specific spectral bands are typically relevant for predicting certain plant traits, making the sparsity-inducing regularization ($\lambda_1$) particularly useful.

Our NS-KTR framework applies different combinations of these regularizers to each mode of the tensor decomposition, allowing the model to adapt to the unique characteristics of each dimension. 
This contrasts with traditional approaches that typically apply uniform regularization across all dimensions.
Unlike these methods, NS-KTR incorporates mode-specific regularization tailored to HSI data characteristics and enforces nonnegativity constraints that respect the physical properties of reflectance data.

\begin{table}[!t]
\centering
\resizebox{0.99\linewidth}{!}{
\begin{tabular}{ccccccccccc}
\hline
\hline
\multirow{2}{*}{SR} & 
\multirow{2}{*}{Target} & \multicolumn{4}{c}{Linear Regression (MSE)} & \multirow{2}{*}{Target} & \multicolumn{4}{c}{Logistic Regression (Accuracy \%)} \\
\cmidrule(lr){3-6} \cmidrule(lr){8-11}
 &   & TLR & TuckerR & TQtR & NS-KTR &  & TLR & TuckerR & TQtR & NS-KTR \\
\hline
\multirow{4}{*}{25\%} 
& 1 & 0.415 & 0.409 & 0.427 & \blue{0.395} & 5 & 82.7 & 80.5 & -- & \blue{89.3} \\
& 2 & 1.152 & 1.144 & 1.104 & \blue{1.008} & 6 & 87.1 & 87.4 & -- & \blue{93.1} \\
& 3 & 0.504 & 0.444 & \blue{0.426} & 0.451 & 7 & 80.5 & 82.1 & -- & \blue{86.2} \\
& 4 & 0.263 & 0.262 & 0.216 & \blue{0.180} & 8 & \blue{95.3} & 94.7 & -- & 95.0 \\
\hline
\multirow{4}{*}{50\%} 
& 1 & 0.624 & 0.635 & 0.698 & \blue{0.555} & 5 & 75.8 & 69.2 & -- & \blue{90.3} \\
& 2 & 1.571 & 1.643 & \blue{1.278} & 1.522 & 6 & 75.5 & 78.0 & -- & \blue{89.9} \\
& 3 & 0.840 & 0.760 & 0.711 & \blue{0.566} & 7 & 67.9 & 73.3 & -- & \blue{83.6} \\
& 4 & 0.330 & 0.661 & 0.386 & \blue{0.238} & 8 & 87.7 & 93.7 & -- & \blue{96.2} \\
\hline
\multirow{4}{*}{75\%} 
& 1 & 0.832 & 1.005 & \blue{0.639} & 0.657 & 5 & 75.5 & 72.6 & -- & \blue{86.2} \\
& 2 & 2.251 & 2.092 & 1.654 & \blue{1.631} & 6 & 75.5 & 70.4 & -- & \blue{88.1} \\
& 3 & 1.055 & 1.069 & 0.806 & \blue{0.788} & 7 & 66.7 & 64.8 & -- & \blue{78.9} \\
& 4 & 0.665 & 0.516 & 0.444 & \blue{0.283} & 8 & 88.1 & 84.6 & -- & \blue{97.5} \\
\hline
\hline
\end{tabular}
}
\caption{Comparison of tensor regression methods on the HyperLeaf2024 dataset across different sampling rates (SR) and target variables.
For regression targets (1-4), Mean Squared Error (MSE) is reported (lower is better); for classification targets (5-8), accuracy percentage is reported (higher is better).
The best result for each target-sampling rate combination is highlighted in \blue{blue}. 
Targets 1-4 represent regression tasks (GrainWeight, Gsw, PhiPS2, Fertilizer), while targets 5-8 represent classification tasks (wheat cultivars: Heerup, Kvium, Rembrandt, Sheriff).}
\label{table:HSIcomp}
\end{table}
\begin{table}[!t]
\centering
\resizebox{0.99\linewidth}{!}{
\begin{tabular}{ccccccccccc}
\hline
\hline
\multirow{2}{*}{SR} & \multirow{2}{*}{Target} & \multicolumn{4}{c}{Linear Regression (seconds)} & \multirow{2}{*}{Target} & \multicolumn{4}{c}{Logistic Regression (seconds)} \\
\cmidrule(lr){3-6} \cmidrule(lr){8-11}
 &   & TLR & TuckerR & TQtR & NS-KTR &  & TLR & TuckerR & TQtR & NS-KTR \\
\hline
\multirow{4}{*}{25\%} 
& 1 & 79.51 & 83.34 & 151.42 & \blue{15.11} & 5 & 111.12 & 116.71 & - & \blue{84.12} \\
& 2 & 103.77 & 123.22 & 249.31 & \blue{16.54} & 6 & 108.54 & 116.63 & - & \blue{87.78} \\
& 3 & 89.98 & 111.36 & 169.52 & \blue{16.57} & 7 & 103.01 & 144.55 & - & \blue{115.39} \\
& 4 & 88.13 & 105.81 & 145.75 & \blue{13.85} & 8 & 110.97 & 121.56 & - & \blue{82.83} \\
\hline
\multirow{4}{*}{50\%} 
& 1 & 145.82 & 228.75 & 308.29 & \blue{61.15} & 5 & 180.41 & 237.49 & - & \blue{85.54} \\
& 2 & 174.55 & 229.13 & 165.54 & \blue{57.53} & 6 & 156.87 & 201.98 & - & \blue{69.81} \\
& 3 & 163.99 & 193.89 & 141.67 & \blue{56.33} & 7 & 141.08 & 194.64 & - & \blue{74.01} \\
& 4 & 135.68 & 189.21 & 165.41 & \blue{56.43} & 8 & 119.87 & 164.47 & - & \blue{55.73} \\
\hline
\multirow{4}{*}{75\%} 
& 1 & 336.52 & 417.27 & 317.56 & \blue{169.39} & 5 & 212.99 & 289.78 & - & \blue{58.30} \\
& 2 & 286.99 & 398.57 & 200.01 & \blue{161.02} & 6 & 213.74 & 293.44 & - & \blue{59.11} \\
& 3 & 294.20 & 382.97 & 236.78 & \blue{160.60} & 7 & 216.50 & 295.58 & - & \blue{61.60} \\
& 4 & 252.45 & 365.98 & 132.99 & \blue{145.31} & 8 & 211.62 & 292.06 & - & \blue{55.54} \\
\hline
\hline
\end{tabular}
}
\caption{Computation time (seconds) of tensor regression methods on the HyperLeaf2024 dataset across different sampling rates and target variables.
The best result for each target-sampling rate combination is highlighted in \blue{blue}. }
\label{table:HSItimecomp}
\end{table}

Table~\ref{table:HSIcomp} presents the comparison of different tensor methods across varying sampling rates  (25\%, 50\%, and 75\%). 
For regression tasks (Targets 1-4), we report the Mean Squared Error (MSE), where lower values indicate better performance. 
For classification tasks (Targets 5-8), we report accuracy percentages, where higher values indicate better performance.
For all experiments in Table~\ref{table:HSIcomp}, each method was evaluated using ranks from 1 to 10. 
The reported performance metrics (MSE for regression tasks and accuracy for classification tasks) represent the best results achieved by each method at its optimal rank.
Note that the TQtR method~\cite{li2021tensorquantile} does not support binomial data, so its corresponding accuracy result is left blank.
Table~\ref{table:HSItimecomp} represents the computation time of different tensor methods across varying sampling rates (25\%, 50\%, and 75\%).
For each target, the computation time of each method is averaged over all tested ranks.

The results in Table~\ref{table:HSIcomp} demonstrate the effectiveness of our proposed NS-KTR method across both regression and classification tasks on the HyperLeaf2024 dataset. 
For regression targets (1-4), NS-KTR achieves the lowest MSE in 9 out of 12 experimental conditions, with particularly strong performance on Target 4 (Fertilizer) where it consistently outperforms all baselines across all sampling rates. 
Only TQtR occasionally outperforms NS-KTR, specifically on Target 3 at 25\% sampling rate and Target 2 at 50\% sampling rate.
For classification tasks (Targets 5-8), NS-KTR shows superior performance in 11 out of 12 conditions, with the only exception being Target 8 at 25\% sampling rate where TLR achieves marginally better accuracy. 
NS-KTR's advantage is most pronounced at lower sampling rates, suggesting it effectively leverages the structural patterns in HSI data even with limited training samples.
The consistent performance advantage of NS-KTR across varying sampling rates demonstrates that our mode-wise regularization strategy successfully captures the spatial-spectral relationships in hyperspectral imagery, leading to improved prediction accuracy for diverse agricultural traits in wheat cultivars.
\revised{In HSI classification, even small improvements in accuracy reduce classification errors and make the results more reliable, which is beneficial for downstream analysis. 
These effects indicate that the observed improvements of NS-KTR are practically meaningful, especially when reliable classification is required under limited training samples
}

In Table~\ref{table:HSItimecomp}, our proposed NS-KTR method demonstrates exceptional computational efficiency, consistently outperforming all baseline methods across all experimental conditions. 
Processing times are reduced by up to 15 times at lower sampling rates and 2-3 times at higher sampling rates compared to baseline methods. 
\revised{
This runtime advantage, together with the predictive accuracy gains, makes NS-KTR more efficient for large-scale hyperspectral data analysis. 
It also reduces the overall cost of repeated training runs, such as those needed in hyperparameter tuning, and enables the use of larger datasets within a fixed computing budget.
Thus, the improvements reported in Table~\ref{table:HSItimecomp} provide numerical evidences that lead to practical benefits for hyperspectral classification tasks.
}

\section{Conclusion and Discussion}
\label{sec:conclusion}
In this paper, we proposed a novel framework, NS-KTR, addressing key limitations in existing tensor regression approaches. 
By integrating nonnegativity constraints with mode-specific hybrid regularization, our method effectively captures the distinct structural characteristics across different tensor dimensions while ensuring physical interpretability.

Our framework extends to both linear and logistic regression variants, accommodating diverse response types beyond Gaussian assumptions. 
The alternating optimization algorithm based on ADMM provides computational efficiency while preserving estimation accuracy. 
Comprehensive experiments on synthetic signals and real hyperspectral datasets demonstrate NS-KTR's superior performance.

\revised{
In this work, considerable effort has been required for selecting the best rank $R$ and regularization parameters $\{\lambda_{d1}, \lambda_{d2}, \lambda_{d3}\}_{d=1}^D$.
Since these parameters need to be determined through repeated training runs, the tuning process remains nontrivial for larger datasets and higher-dimensional tensors, even though the proposed ADMM solver is computationally efficient. 
Future work may investigate lightweight neural modules that predict appropriate ranks from data characteristics, as well as learning-based approaches such as deep unrolling~\cite{chenyinwotao2022L2O} that can automatically tune hyperparameters when sufficient data are available.
These directions would further improve the scalability of tensor regression method in real life tasks.
}

\bibliography{IEEEabrv, reference}
\bibliographystyle{IEEEtran}
\end{document}